\documentclass[sigconf,9pt,authorversion,nonacm]{acmart}
\def\BibTeX{{\rm B\kern-.05em{\sc i\kern-.025em b}\kern-.08emT\kern-.1667em\lower.7ex\hbox{E}\kern-.125emX}}
    
\usepackage{nicefrac}
\usepackage{siunitx}
\usepackage{array,framed}
\usepackage{booktabs}
\usepackage{
  color,
  float,
  epsfig,
  wrapfig,
  graphics,
  graphicx,
  subcaption
}
\usepackage{setspace}
\usepackage{latexsym,fancyhdr,url}
\usepackage{enumerate}
\usepackage[linesnumbered,ruled,vlined]{algorithm2e} 
\usepackage{algpseudocode}
\usepackage{graphics}
\usepackage{xparse} 
\usepackage{xspace}
\usepackage{multirow}
\usepackage{csvsimple}
\usepackage{balance}
\usepackage{bbm}
\usepackage[font=small]{caption}
\usepackage{enumitem}

\newcommand{\name}{DeepStream}

\newcommand{\beitong}[1]{\textcolor{orange}{{beitong: #1}}}

\begin{document}

\title{DeepStream: Bandwidth Efficient Multi-Camera Video Streaming for Deep Learning  Analytics
}

\author{Hongpeng Guo}
\email{hg5@illinois.edu}
\affiliation{%
  \institution{UIUC}
}

\author{Beitong Tian}
\email{beitong2@illinois.edu}
\affiliation{%
  \institution{UIUC}
}

\author{Zhe Yang}
\email{zheyang3@illinois.edu}
\affiliation{%
  \institution{UIUC}
}

\author{Bo Chen}
\email{boc2@illinois.edu}
\affiliation{%
  \institution{UIUC}
}

\author{Qian Zhou}
\email{qianz@illinois.edu}
\affiliation{%
  \institution{UIUC}
}

\author{Shengzhong Liu}
\email{sl29@illinois.edu}
\affiliation{%
  \institution{UIUC}
}

\author{Klara Nahrstedt}
\email{klara@illinois.edu}
\affiliation{%
  \institution{UIUC}
}

\author{Claudiu Danilov }
\email{claudiu.b.danilov@boeing.com}
\affiliation{%
  \institution{Boeing Research}
}

\begin{abstract}
Deep learning video analytic systems process live video feeds from multiple cameras with computer vision models deployed on edge or cloud. To optimize utility for these systems, which usually corresponds to query accuracy, efficient bandwidth management for the cameras competing for the fluctuating network resources is crucial.
We propose \name{}, a bandwidth efficient multi-camera video streaming system for deep learning video analytics. \name{} addresses the challenge of limited and fluctuating bandwidth resources by offering several tailored solutions. We design a novel Regions of Interest detection (ROIDet) algorithm which can run in real time on resource constraint devices, such as Raspberry Pis, to remove spatial redundancy in video frames and reduce the amount of data to be transmitted. We also propose a content-aware bandwidth optimization framework and an Elastic Transmission Mechanism that exploits correlations among video contents.  We implement \name{} on Raspberry Pis and a desktop computer. Evaluations on real-world datasets show that \name{}'s ROIDet algorithm saves up to 54\% bandwidth with less than 1\% accuracy drop. Additionally, \name{} improves utility by up to 23\% compared to baselines under the same bandwidth conditions. 
\end{abstract}

\maketitle

\section{Introduction}

Deep learning video analytic systems are widely deployed in various areas, such as stores, traffic intersections, and parking lots for customer experience \cite{amazon}, traffic monitoring \cite{canel2019scaling}, safety surveillance \cite{newart}, and more. These systems perform computer vision tasks on live video streams using classification \cite{he2016deep, howard2017mobilenets}, object detection \cite{redmon2016you}, or object tracking \cite{nam2016learning, Wojke2017simple} to detect objects or activities.

Analytic accuracy is crucial in these applications. However, cameras have limited on-device computing power \cite{li2020reducto} and cannot execute large models on live video feeds. Therefore, recent works propose to offload tasks to a cloud or edge server for execution \cite{zhang2018awstream,du2020server,ran2018deepdecision}. However, transmitting high-quality video feeds to remote servers is bandwidth-intensive, especially when multiple cameras compete for limited bandwidth resources.

In this work, we aim to achieve high-accuracy queries when streaming videos from a fleet of co-located cameras under limited bandwidth conditions for object detection, which is an important task in computer vision and a key component in other downstream tasks such as object tracking. Recent works tackle this issue by reducing spatial \cite{du2020server, zhang2021elf, zeng2020distream, liu2022multi} and temporal video redundancy \cite{li2020reducto, kang2017noscope, canel2019scaling} or making bandwidth allocation decisions \cite{wang2020joint, zhang2018awstream, zhang2015design}. However, they either rely on heavy-weighted computer vision models to detect  redundancy in video frames or do not fully leverage video redundancy in bandwidth optimization. We propose \name{}, a novel multi-camera video streaming system that integrates   video spatial redundancy removal and bandwidth allocation to optimize the  detection utility, which corresponds to detection accuracy, given limited and fluctuating bandwidth conditions.

\name{} has three components to achieve this goal. First, we remove spatial redundancy in video frames to reduce the amount of data that needs to be transmitted. We design a novel  Regions of Interest detection algorithm that can run in real time on resource-constrained camera devices and does not require feedback information from the server. Second, we propose a framework to maximize object detection utility by formulating multi-camera bandwidth allocation as a content-aware optimization problem. We use offline profiling to discover the relationship between detection utility and video content, and design a content-aware dynamic programming based bandwidth allocation algorithm. Third, we propose an Elastic Transmission Mechanism that adapts to spatial and temporal correlations among video contents to improve detection utiltiy.
In summary, we make the following three contributions in this paper:

\begin{enumerate}[leftmargin=*,label= ({\textbf{\arabic*}})]
    \item We design a novel Regions of Interest detection algorithm that  removes spatial redundancy in videos. This is the \emph{first} accurate ROIs detection algorithm which runs in real time on resource constraint devices and does not require feedback ROIs information from the server.
    \item We propose a content-aware optimization algorithm for multi-camera bandwidth allocation, utilizing dynamic programming. We leverage spatial and temporal correlations in video contents to improve detection utility.
    \item We implement \name{} camera-side processing module on Raspberry Pis and server-side processing on a desktop-level computer. Real-world evaluations show that ROIDet saves up to 54\% bandwidth with less than 1\% accuracy drop, and overall \name{} enhances utility by up to 23\% compared to baselines under the same bandwidth conditions. 
\end{enumerate} 

\section{Motivation and Preliminaries}

In this section, we present the special characteristics of video streaming for deep learning analytic applications, then we discuss the design scope of \name{}.

\subsection{Motivation and Background}
This work focuses on deep learning video analytics for co-located cameras. These systems stream live video feeds to an edge server or cloud server for processing with computer vision models. Besides addressing limited and fluctuating network bandwidth, special characteristics need to be considered.

\noindent\textbf{Videos have a large portion of spatial redundancy.} Since we target video streaming for analytics, computer vision models are executed on video frames to classify, detect, or track objects on the frames.  We analyze the videos discussed in Section~\ref{subsec:data} to examine what size of frame area is considered foreground and contains objects. Only these frame areas contain objects of interest.
We find in this dataset more than 80\% of the frames  have less than 50\% task-related areas. Removing the spatial redundancy in video frames can result in major bandwidth savings.

\noindent\textbf{Constraint on-camera computing power.} Cameras deployed in places such as traffic crossings do not have the same computing power as cloud or  desktop-level computing devices \cite{li2020reducto}. A typical computing device \cite{rpi} used for live video streaming has only a 4-core 1.5GHz CPU, 2GB RAM, and no dedicated GPU resources.  Heavy-weighted processing models \cite{zhang2021elf} are not able to run on device, making it prohibitive to use complex models to remove spatial or temporal redundancy on device.

\noindent\textbf{Cameras are co-located and there are spatial-temporal correlations in video contents.} Since we consider scenarios such as traffic monitoring, usually a fleet of cameras are deployed in the same area capturing the same set of objects simultaneously or sequentially. Because of this characteristic, video content streamed from co-located cameras is also synchronously fluctuating. 

\subsection{Design Scope of \name{}}
\label{sec:scope}

Considering network bandwidth conditions and the aforementioned characteristics, we set the design space for \name{}.

\noindent\textbf{Video compression.} Video codecs such as H.264 \cite{telecom2003advanced} provide the benefit of reducing redundancy between and within video frames to decrease the amount of data to transmit. \name{} integrates video codecs into its design. Specifically,  we divide time into time slots with length $T$. Every camera $i$ gathers $N$ frames which arrive in each time slot, compresses these frames into a video segment with a frame rate to be $N/T$, and  transmits the segment to the server for processing. The frames can be compressed into segments of different bitrates $b_i\in B$ and resolutions $r_i\in R$. We study the impact that video compression poses on query utility  in Section~\ref{subsec:utility}. It is worth mentioning that, since video compression is applied, some redundancy in videos, especially temporal redundancy, is naturally reduced (see Section~\ref{ssec:vs}). Therefore, we only focus on Regions of Interest based spatial redundancy removal in videos.

\noindent\textbf{On camera video pre-processing.} To reduce the spatial redundancy in video frames as mentioned previously, we enable on camera video processing to crop out task irrelevant areas. The limited computing power of cameras prohibits us from using large computer vision models. So we limit ourselves to use light-weighted algorithms to pre-process the videos for transmission. 
In addition, \name{} targets static cameras such as traffic cameras but can be easily adapted to incorporate non-static cameras.

\noindent\textbf{\name{} target.}  
\name{} is a camera-edge collaborative framework. The goal of \name{} is to maximize the utility for co-located cameras when videos are transmitted to a central edge server for processing with computer vision models over constraint bandwidth networks.
Specifically, in this work, we focus on convolutional neural network based object detection tasks (\emph{eg.} Yolo) as object detection is widely used today and is a prerequisite for downstream tasks such as object tracking \cite{Wojke2017simple}.  Since videos are consumed by machine learning models, we use detection accuracy, and F1 score, as utility metrics. F1 score measures detection accuracy by computing the  harmonic mean of the precision and recall. In this way, during each time slot, the utility of a camera is the detection F1 score on the video segment that it streams to the edge. And the total utility in each time slot is the weighted sum of all cameras' utility, where weights are user-defined values.

\section{\name{} System Overview}

 \begin{figure}
  	\centering
  	\includegraphics[width=1\linewidth]{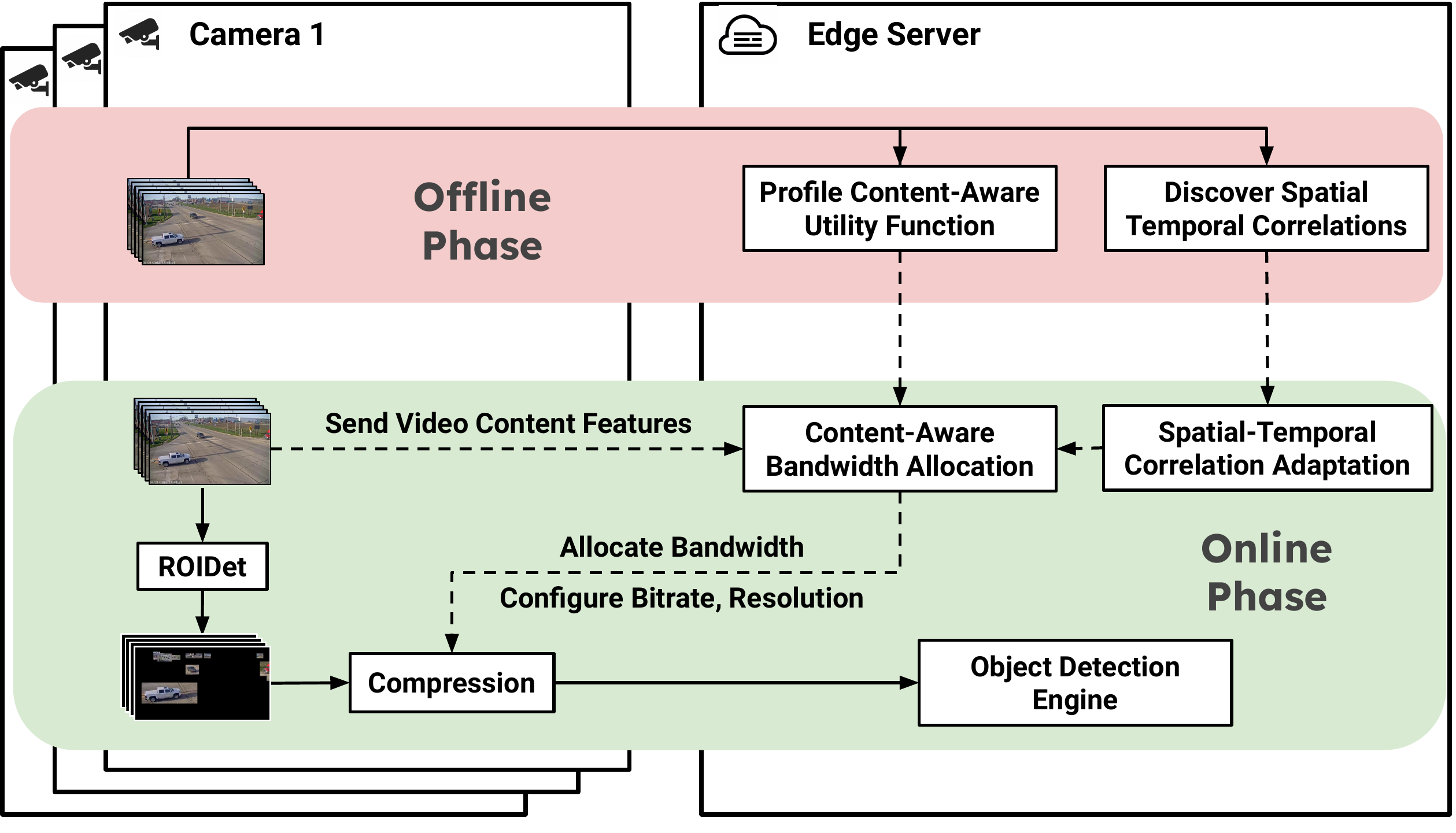}
  	\vspace{1mm}
	\caption{Overview of \name{}. Dashed and solid arrows represent control plane and data plane communication, respectively.}
	\label{fig:overview}
\end{figure}

\name{} operates in two phases as shown in Figure~\ref{fig:overview}. (1) In the offline profiling phase, \name{} profiles how detection utility varies over different video content features and video compression configurations, including bitrate and resolution. It also discovers spatial and temporal correlations between video contents to be used in bandwidth management. (2) In the online phase, \name{} has a camera side 
 \textbf{ROIDet Module}, which runs in real time on low end computing devices to discover Regions of Interest on video frames. The ROIDet Module crops video frames to remove irrelevant regions which don't contain objects. Based on the allocated bandwidth and bitrate received from the server, the camera compresses video frames into a video segment and sends the segment to the edge server for processing with a sending rate decided by the allocated bandwidth. To decide bandwidth allocations for cameras, \name{} has a content-aware \textbf{Bandwidth Allocation Module}, which optimizes detection utility given current video content features and the utility profile obtained from the offline phase. To assist bandwidth allocations, we design a \textbf{Spatial and Temporal Adaptation Module} to adapt to correlated video content variations among cameras and network bandwidth variations.

\section{\name{} Camera}
\label{sec:cam}

The main task of \name{} camera side processing  is to detect regions on image frames that contain objects, to compress these frames into video segments according to bitrate and resolution configurations determined on the server, and to send these video segments to the server. In this section, we present the Regions of Interest detection and cropping technique in detail. 

 \begin{figure}
  	\centering
  	\includegraphics[width=0.8\linewidth]{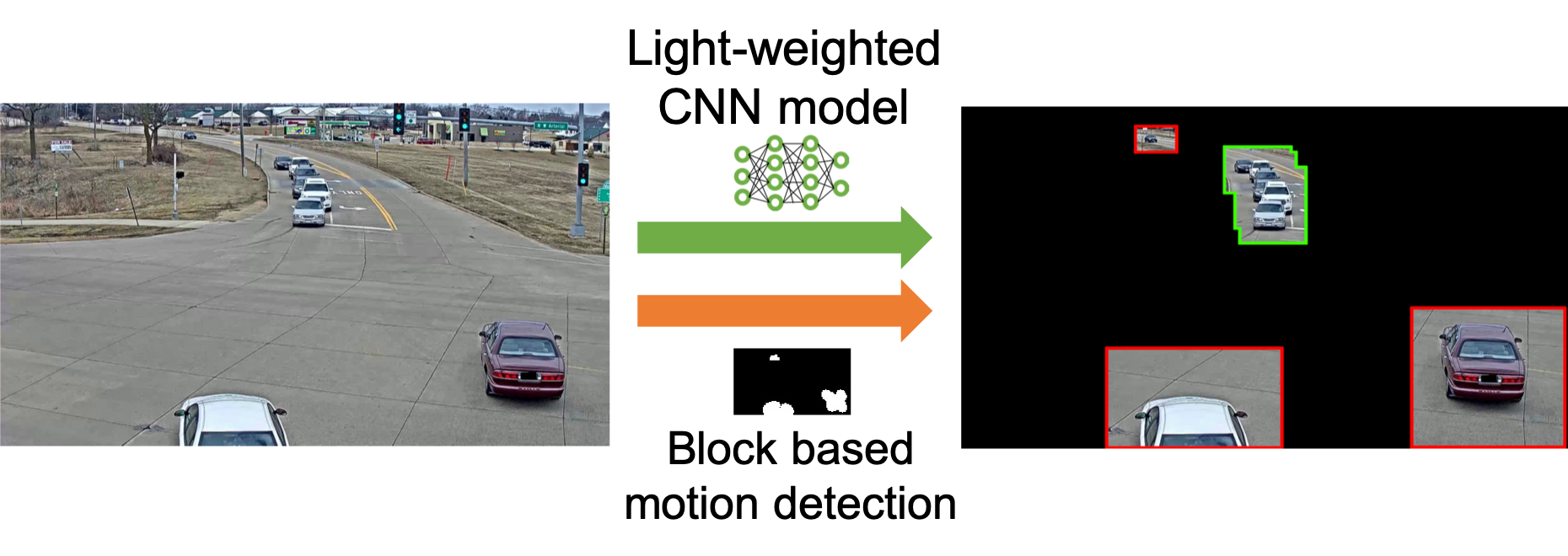}
	\caption{ROIDet workflow. We use a light-weighted object detection model to detect stationary objects (in green bounding boxes), and design a block based motion detection algorithm to detect moving objects (in red bounding boxes).}
	\label{fig:roi_det}
	\vspace{-5mm}
\end{figure}
The Regions of Interest Detection (ROIDet) component detects regions within a group of frames captured from static cameras which potentially contain objects, so that cameras only need to encode and send these relevant regions to the  server in order to save network bandwidth. 
There are some works proposing to encode and send only Regions of Interest at the camera side, but these works either rely on expensive object detectors deployed on the device to detect potential regions containing objects \cite{zhang2021elf}, or use detection feedback from the server to detect relevant regions \cite{du2020server}. The first approach relies on hardware with abundant GPU computing resources at the camera side; the second method needs to send the server low quality videos first which consumes extra bandwidth resources, and it also needs to wait for server feedback before the camera can start to encode videos.

We design a novel Regions of Interest detection algorithm, shown in Figure~\ref{fig:roi_det}, for static cameras such as traffic cameras, which are able to run in real time on low end computing devices such as Raspberry Pis.
Our proposed method uses a light-weighted object detector and image low level processing algorithms, and it can achieve $~15fps$ detection for 1080p videos and $~30fps$ for 720p videos on Raspberry Pis.
The rationale driving the design of this algorithm is as follows. 

Objects in a video segment are either moving or stationary. 
For \textbf{moving objects}, we can use the differences in frames' low level features between consecutive frames (\name{} uses edge differences) to detect where motion exists between  frames, by examining the distribution of low level feature changes on the consecutive frames. In this way, we are able to detect the estimated locations of moving objects. As a matter of fact, background subtraction algorithms \cite{zivkovic2006efficient} are proposed to subtract foreground moving objects from background. However, they typically require higher computing demands.
 To detect \textbf{stationary objects}, traditional image processing techniques usually perform poorly. Therefore, we utilize a convolutional neural network based object detector to find objects that are not moving. However, object detectors such as Yolo \cite{redmon2016you} are resource hungry and not able to achieve real time detection on low end devices. Therefore, we utilize a light-weighted version of Yolo \cite{chen_2021_5241425} which is able to achieve fast detection on low end devices with low resolutions. To further reduce the likelihood of missing objects on a frame due to low quality model and low resolution video frames, we configure our object detector to run at a low confidence threshold. Since we need light-weighted Yolo to detect only stationary objects in a video segment, we run the object detector once per video segment.

 \begin{algorithm}[t]
\SetAlgoLined

\KwIn{Video segment consisting of $N$ frames $G=\{g^{(1)},g^{(2)},...,g^{(N)}\}$}
\KwOut{Bounding boxes $B$ corresponding to Regions of Interest in $G$}
 $B_{1}=objectDetector(g^{(1)})$\;
 \For{frame  $g^{(i)}$}
 {$e^{(i)} = cannyEdgeDetection(g^{(i)})$\;
 $\Delta e^{(i)} = e^{(i)} - e^{(i-1)}$
 
 Partition $\Delta e^{(i)}$ into $M\times N$ blocks\;
 Initialize empty matrix $D_{M\times N}$\;
 \For{each block $\Delta e^{(i)}_{m,n}$}
 {
 $D_{m,n}\gets\mathbbm{1}(\sum\Delta e^{(i)}_{m,n} > threshold)$
 }
 }
 $B_{2}=getConnectedComponents(D)$\;
 \Return $B_{1}\bigcup B_{2}$

 \caption{ROIs Detection Algorithm}
 \label{al:roialgo}
\end{algorithm}

We describe ROIDet in Algorithm.~\ref{al:roialgo}. The input to this algorithm is a series of video frames that a camera captures during the past time slot, and the output is a set of bounding boxes which contain all the objects in a video segment. 
First, we run a light-weighted CNN based object detector \cite{chen_2021_5241425} to detect all stationary objects in the video segment. Then on line 3 to 4, we calculate the differences of image edges between a frame and its previous frame, and obtain an edge difference matrix where $1$ represents a change of edges at the corresponding pixel location and $0$ means no edge changes. We partition the edge difference matrix into $M\times N$ blocks. From line 6 to line 9, for each block, we sum up all values in each block and compare the sum with a predefined threshold value. If the sum of the edge difference values in a block is greater than the threshold, we think there is motion in the block; vice versa. We obtain a binary matrix $D_{M\times N}$. Then on line 11, we perform the Spaghetti Labeling Algorithm \cite{bolelli2019spaghetti} on $D_{M\times N}$ to find rectangular bounding boxes for all the connected components (a connected component is a set of $1$s in a matrix that are connected to each other) in $D_{M\times N}$. We merge the bounding boxes found with the light-weighted CNN based object detector and our block based moving object detector to obtain the bounding boxes for a video segment.

Once a camera obtains detections from ROIDet, it sends detection confidence and ROIs to the server as video content features, and starts to compress and transmit the video frames once it receives configuration and bandwidth allocation from the server.

\section{\name{} Server}
\label{sec:ser}

\name{} is a collaborative processing framework between multiple cameras and a server. The server side processing  consists of $3$ components, namely, offline Utility Function Profiling, Bandwidth Allocation, and Spatial-Temporal Correlation Adaptation.

\subsection{Utility Function Profiling}
\label{subsec:utility}
As we mention in Section~\ref{sec:scope}, we use object detection accuracy as utility, and aim to maximize the total utility across all cameras in each time slot, which corresponds to the weighted sum of cameras' detection accuracies. We systematically evaluate different factors which potentially affect detection accuracy, and select 4 significant factors which fit well in the design of \name{}.


As we already know, object detection accuracy is a function of video bitrate and resolution --- the larger bitrate a video segment gets, the less it is compressed, and therefore the higher detection accuracy is; resolution is related to object sizes on a video frame, and therefore affects detection accuracy too. Meanwhile, detection accuracy is also a function of video content. As we discuss in  Section \ref{sec:cam}, we crop out the video background and only compress and transmit regions which potentially contain objects. These Regions of Interest (ROI) vary over time and are unique for different cameras. Given the same compression bitrate, the larger the ROI is, the fewer bytes the video codec can leverage in compressing a pixel. Therefore, detection accuracy is a function of the ROI size in a video segment. Moreover, detection accuracy is also highly related to how difficult it is for a machine learning model to successfully detect objects in a video segment. Several factors affect the performance of the object detection model, such as object sizes and object angles. Before executing an object detection model on a video segment on edge, we cannot know exactly how well the model performs on that segment. Luckily, executing a lighted-weighted object detection model on camera, as discussed in Section \ref{sec:cam}, gives us some hints on detection difficulty. Although on camera we use a reduced version of the server side model, these two models share similar structures and therefore react similarly to different object sizes and angles. We use the detection confidence score as an indicator of detection difficulty.

In this way, in \name{}, we set our utility, which is detection accuracy, to be a function of video content, which corresponds to the ratio of the size of Regions of Interest over the size of the full frame of a video segment $a\in [0,1]$, and the average on camera detection confidence  $c\in [0,1]$. These data are given by the ROIDet Module discussed in Section~\ref{sec:cam}. Detection accuracy is also a function of bitrate $b\in B$ and resolution $r\in R$. So for each camera $i$, we need to profile an accuracy function $\hat{\alpha_i}=f_i(a_i,c_i,b_i,r_i)$ to get predicted accuracy $\hat{\alpha_i}$. When a camera is first deployed, we send uncropped, highest bitrate, and highest resolution frames to the edge server for profiling, in order to learn the accuracy function. We use a two-layer fully connected neural network as the regression model as the accuracy function. The inputs to this neural network are $a_i,c_i,b_i,r_i$. And the output is the predicted accuracy $\hat{\alpha_i}$.

\subsection{Bandwidth Allocation}

At each time slot $t$, each camera sends its video content features $a_i^{(t)}$ and $c_i^{(t)}$ to the server, and the server uses these features to predict their accuracy functions $\hat{\alpha_i}$ and optimizes the global accuracy over all cameras:
\begin{equation}
\vspace{-1mm}
\label{equ:opt}
\begin{gathered}
    \textstyle\max \sum_i \lambda_i \hat{\alpha_i}^{(t)} = \max \sum_i \lambda_i f_i(a_i^{(t)},c_i^{(t)},b_i^{(t)},r_i^{(t)}), \\
    b_i^{(t)}\in B, \\ 
    r_i^{(t)}\in R, \\
    \textstyle\sum_i b_i^{(t)} \leq W^{(t)},
  \end{gathered}  
  \vspace{-2mm}
\end{equation}
where $\hat{\alpha_i}^{(t)}$ is the predicted accuracy of camera $i$ in time slot $t$ given ROIs area ratio $a_i^{(t)}$, detection confidence $c_i^{(t)}$, compression bitrate $b_i^{(t)}$ and resolution $r_i^{(t)}$. $\lambda_i$ is the utility weight of camera $i$. The first and second constraints list all possible bitrate and resolution options that camera $i$ can take\footnote{Bitrate can also be continuous. In that way this problem is solved with heuristics based algorithm.}. The third constraint expresses that selected video bitrates across all cameras cannot exceed current bandwidth $W^{(t)}$.

The time complexity for exhaustive search of problem (\ref{equ:opt}) in each time slot is exponential with the number of cameras. However, this problem can be reduced to a special case of knapsack problem and therefore be solved within pseudo-polynomial time. In \name{}, we further reduce the variable space by neglecting $a_i$ and $c_i$ in the accuracy function since they are fixed in the optimization process. Finally, we solve this optimization problem using dynamic programming algorithm which achieves time complexity of $O(|I||B| \frac{|W|}{d})$ ,where $|I|$ is the number of cameras, $|W|$ is the total bandwidth constraint, and $d$ is the greatest common factor of all possible bitrate configurations.

\subsection{Adaptation to Spatial-Temporal Correlations}
\label{sec:spatem}

Since \name{} targets network scheduling for a fleet of co-located cameras, spatial-temporal correlations exist among the video contents \cite{jain2020spatula}. Specifically, the same objects may enter or exit the fields of view in multiple cameras at approximately the same time.  
As we perform Regions of Interest based frame cropping in Section~\ref{sec:cam}, the areas of ROIs are therefore correlated across the cameras.
For large ROI areas, high bandwidth is required to maintain high compression quality and detection accuracy. Small ROIs do not require high bandwidth. 
Moreover, network bandwidth conditions vary over time. When more bandwidth is needed to encode larger ROIs, this requirement may not be satisfied by the current network conditions. Based on these insights, we propose a novel elastic transmission mechanism for \name{}. 

\subsubsection{Thresholds Selection}
 \name{} performs an analysis of video contents to obtain two thresholds --- a threshold on the total area of ROIs, and a threshold on the current bandwidth condition. If the total area of ROIs is larger than its threshold and the current network bandwidth is smaller than its threshold, \name{} borrows transmission time from future time slots.

(a) To obtain the threshold on area of Regions of Interest, \name{} keeps a moving average of the area of Regions of Interest $\hat{a}^{(t)}=\alpha a^{(t)}+(1-\alpha)\hat{a}^{(t-1)}$ where $a$ represents the total area of Regions of Interest across all cameras, and $\sigma_a$ denotes its standard deviation. The threshold can therefore be expressed as 
    $\tau_a = \hat{a}^{(t)} + \gamma_a \sigma_a$, 
where $\gamma_a$ is a parameter controlling the aggressiveness of elastic transmission. This threshold selection process happens both offline and online.
(b) To obtain the threshold on bandwidth conditions, \name{} first selects the configuration that achieves the highest detection accuracy for each bitrate option.
For each bitrate option $b$ and for each camera $i$, \name{} calculates the standard deviation of accuracy differences between encoding in bitrate $b_i$ and encoding in the highest possible bitrate $b_M$ for all video segments in the profiling set. When the standard deviation is above an empirically preset value $\sigma_{high}$, \name{} selects the sum of corresponding bitrates $\tau_{wl}=\sum_i b_i$ as a threshold of demanding more transmission time. When the standard deviation is lower than $\sigma_{low}$, \name{} selects the corresponding bitrate sum as a threshold $\tau_{wh}$ of being able to give back transmission time. This process happens offline.

\subsubsection{Transmission Adjustment}


When $a^{(t)}>\tau_a$ and $W^{(t)}<\tau_{wl}$, \name{} borrows extra transmission time by delaying the start time of the next time slot, and uses the additional time to transmit data for the current time slot. The amount of data transmitted during this borrowed time is $D=\gamma_{wl}(\tau_{wl}-W^{(t)})T$, where $\gamma_{wl}$ controls the level of aggressiveness, and $T$ is time slot length. A budget limit is set for $D$, and \name{} stops borrowing time when the extra transmitted data reaches this budget.
On the other hand, when $W^{(t)}\geq\tau_{wh}$, DeepStream replenishes the budget by finishing its time slots earlier. This mechanism is integrated into Bandwidth Allocation by modifying the third constraint in Equation~\ref{equ:opt} to be $\sum_i b_i^{(t)} T\leq W^{(t)}T+D$.

\section{Implementation}

We implement \name{} on Raspberry Pis and a desktop-level computer with Python. 

\noindent\textbf{Hardware Settings}
The camera side processing module of \name{} is implemented on Raspberry Pi 4B, which has a 4-core 1.5GHz CPU and 2GB RAM. The server has an Intel  i7-9700K 8-core CPU, 64GB memory, and a GeForce RTX 2080 Graphics Card with 2944 CUDA cores.
The cameras are connected to a dual-band WiFi router. The server is connected to the router's Ethernet port. 

\noindent\textbf{Network Stack Configurations}
We use UDP for data transmission in the data plane, implementing an ACK-based retransmission mechanism for reliable transmission. In the control plane, which includes video content features sharing and bandwidth allocation assignment, we use ZeroMQ \cite{zmq} based on TCP 
We use the \texttt{police} action of the Linux traffic control tool \texttt{tc} to enforce bottleneck bandwidth constraint on the server side.
To handle available bandwidth probing, we use packet pair probing as used by UDT \cite{gu2007udt}.

\noindent\textbf{Video Processing Tools}
On camera side, we implement the block based moving object detection algorithm with OpenCV, and we use a light-weighted object detection model YOLOv5-Lite \cite{chen_2021_5241425} to detect stationary objects in a video segment. Videos are encoded on camera and decoded at the server side. We use libx264 provided by FFmpeg \cite{ffmpeg} as the video encoder and FFmpeg to decode videos into frames for inference.

\section{Evaluation}
We evaluate the performance of \name{} by answering these questions:
\begin{itemize}
    \item Is ROIDet efficient in terms of saving network bandwidth? Does this cropping technique induce accuracy loss?
    \item How well does the Elastic Transmission Mechanism handle bandwidth and video content variations? 
    \item What is the end-to-end performance of \name{} in terms of accuracy and latency?
\end{itemize}

\subsection{Experiment Dataset}
\label{subsec:data}
We use a public video dataset AI City Challenge \cite{Naphade21AIC21}. In this dataset, we use video feeds from five traffic cameras which are co-located and monitoring a traffic intersection. The resolution of the first four videos is $1080\times1920$ and the resolution of the last video is $960\times 1280$. All videos have $10$ frames per second. In our experiments, videos are separated into video segments of $1$ second with each segment consisting of $10$ frames. We encode each video segment into bitrates ranging from $50Kbps$ to $1000Kbps$. 
For each bitrate, we encode video segments into three different resolutions. 
We use the first $80s$ of each video to train the accuracy function models and to obtain thresholds for the Elastic Transmission Mechanism and use the rest $120s$ of each video to evaluate \name{}'s performance.

We use bandwidth traces published by the FCC \cite{fcc}. We compile 3 traces from the dataset and classify them into \emph{low}, \emph{medium}, and \emph{high} bandwidth conditions. The low, medium and high bandwidth traces have means and standard deviations of 521 Kbps/230 Kbps, 1134 Kbps/499 Kbps, and 2305 Kbps/1397 Kbps, respectively.
We set the network propagation delay to be $20ms$.

\subsection{\name{} vs. State-of-the-art Systems}
\label{ssec:vs}

We compare \name{} with state-of-the-art video analytics systems\footnote{It is worth mentioning that we are aware of DDS \cite{du2020server}, which first transmits a low-quality video stream to the server to determine ROI, and feedback these regions to the camera to transmit a higher-quality cropped video. Since they use a two-round transmission protocol, it is difficult to directly compare their performance with ours with regard to utility. However, we do evaluate the efficacy of our ROIs cropping technique  in Section~\ref{sec:roi}.} in delivering high utility detection under constrained bandwidth conditions:
    
\noindent\textbf{Reducto} \cite{li2020reducto} reduces frames transmitted by filtering frames on-camera using low-level image features and compresses filtered frames for transmission. While Reducto operates on single-camera streams, we allocate bandwidth fairly among cameras to decide each video stream's bitrate.
    
\noindent\textbf{JCAB} \cite{wang2020joint} optimizes configuration adaptation and bandwidth allocation for multiple cameras by solving an optimization problem to maximize global utility. Their utility function is expressed as a function of video resolution. To adapt JCAB to our framework, we enhance their utility function to also consider video bitrate, ensuring a fair comparison.
     
\noindent\textbf{\name{} without Elastic Transmission.} We disable Elastic Transmission in DeepStream to study the efficacy of the mechanism and compare its performance.

We compare the segment utility achieved by each approach, where segment utility refers to the weighted sum of detection accuracies of all cameras in a time slot. We compute the average segment utility over all segments/time slots. To demonstrate the performance of \name{} under different configurations, we assign different weights $\lambda_i$ to the utility function of different cameras $i$ in Equation~\ref{equ:opt}. First, we give all the cameras the same weights (1.0). Then, we use a uniform distribution to randomly generate
a new set of weights for the cameras. 
The generated weights used in our evaluation are $0.84$, $0.38$, $1.92$, $0.74$, and $0.45$ for cameras $1$ to $5$, respectively. We use the network traces mentioned previously.

\begin{figure}
  \centering
  \begin{subfigure}{0.23\textwidth}
      \includegraphics[width = 1.0\linewidth]{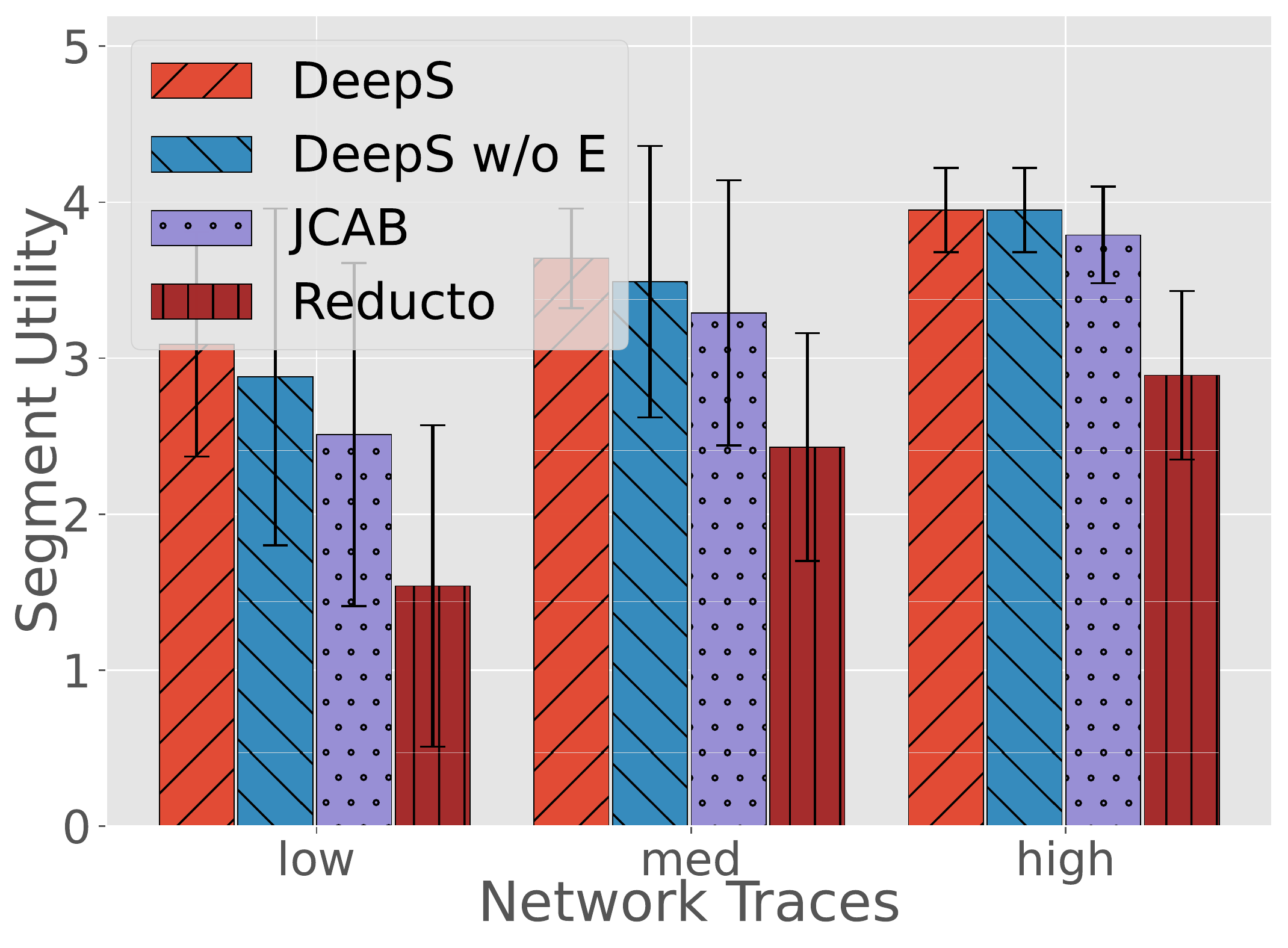}
      \caption{$\lambda$ set 1.}
      \label{fig:eval-memory-server}
  \end{subfigure}
  \hfill
  \begin{subfigure}{0.23\textwidth}
      \includegraphics[width = 1.0\linewidth]{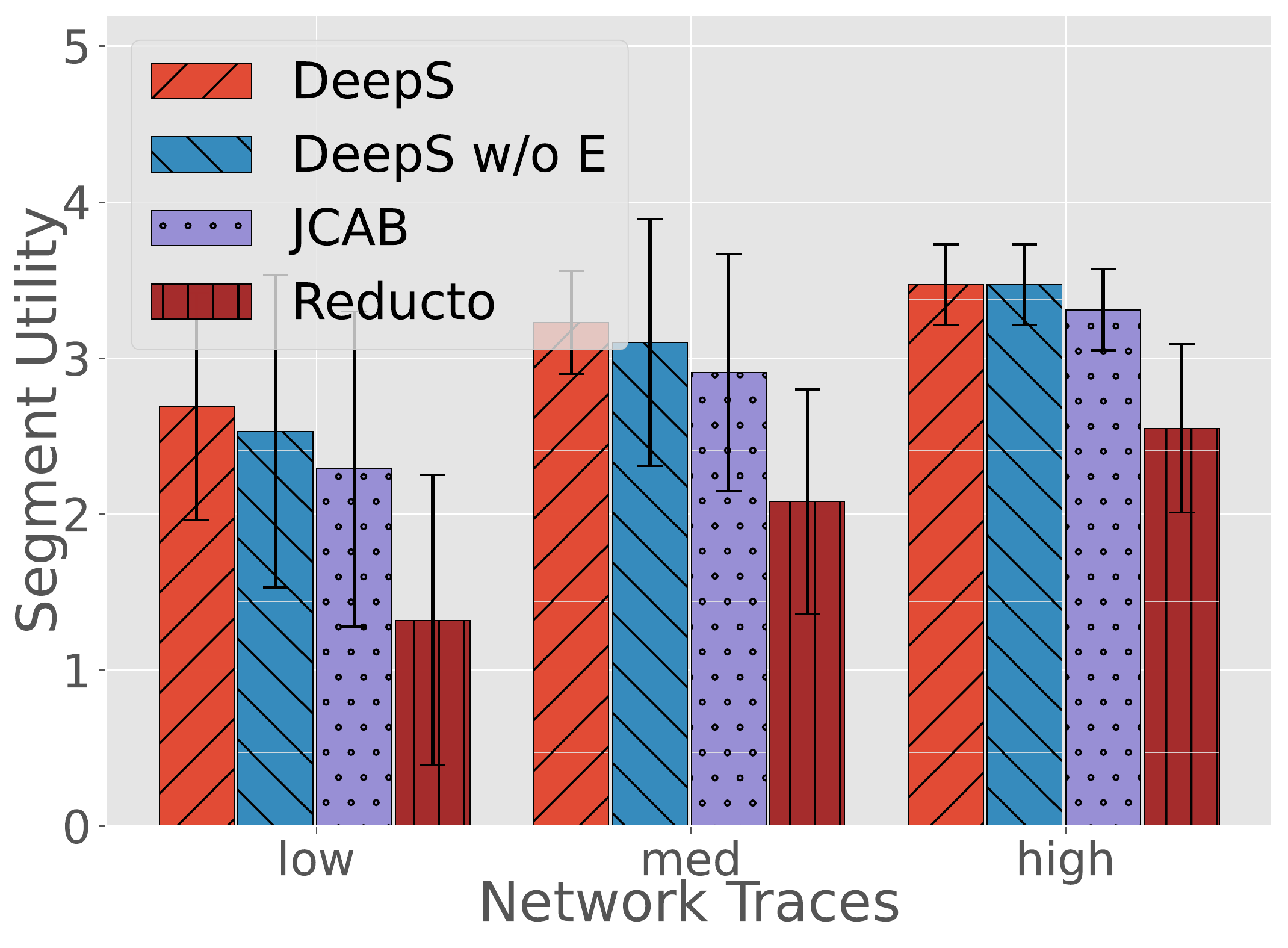}
      \caption{$\lambda$ set 2.}
      \label{fig:eval-memory-jetson}
  \end{subfigure}
  \vspace{-3mm}
  \caption{Compare \name{} against state-of-the-art machine centric streaming  systems with regard to their abilities to achieve high utility under different utility functions and different network bandwidth conditions.}
  \label{fig:overall}
 \end{figure}

Figure~\ref{fig:overall} shows the comparison results of \name{} against other systems. We can see that \name{} outperforms all other systems under all network traces and utility function settings.  We can also observe that the performance difference is the largest when the available bandwidth is low, demonstrating \name{}'s ability to provide high utility transmission under constraint bandwidth conditions. 
Reducto performs poorly compared to other systems because it redundantly filters frames, which video codecs\footnote{In the Reducto\cite{li2020reducto} paper, the authors mention that they use the H.264 standard to compress the filtered frames.} already compress, resulting in suboptimal redundancy removal.

\subsection{Regions of Interest Cropping}
\label{sec:roi}

We evaluate the performance of ROIDet. Specifically, we evaluate (1) under the same bitrate conditions, how much accuracy gain \name{} produces, and (2) whether \name{} losses accuracy compared with non-cropping encoding, when videos are encoded with the highest possible quality.  

\begin{figure}
\vspace{-3mm}
  \centering
  \begin{subfigure}{0.23\textwidth}
      \includegraphics[width = 1.0\linewidth]{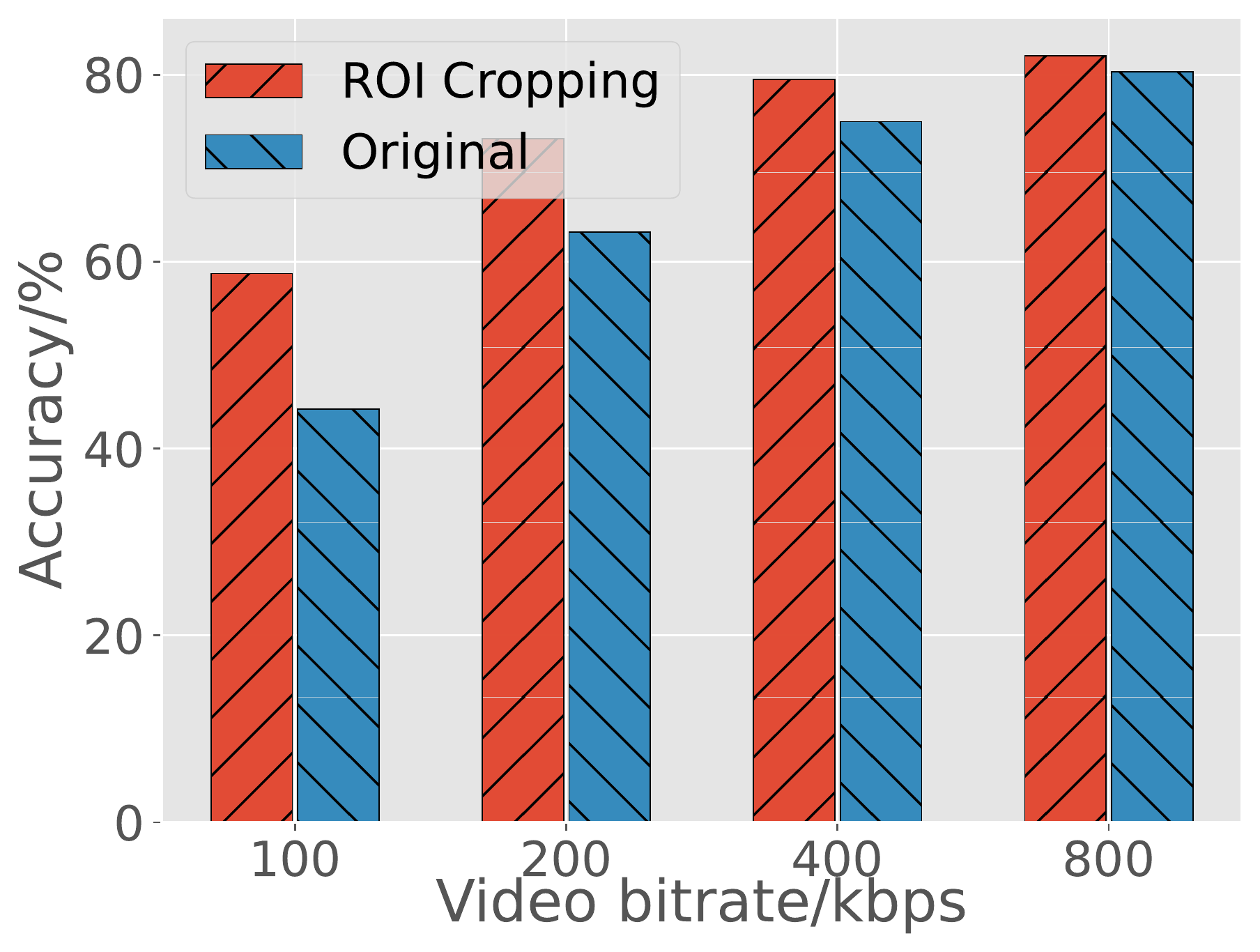}
      \caption{720p.}
      \label{fig:acc-720}
  \end{subfigure}
  \hfill
  \begin{subfigure}{0.23\textwidth}
      \includegraphics[width = 1.0\linewidth]{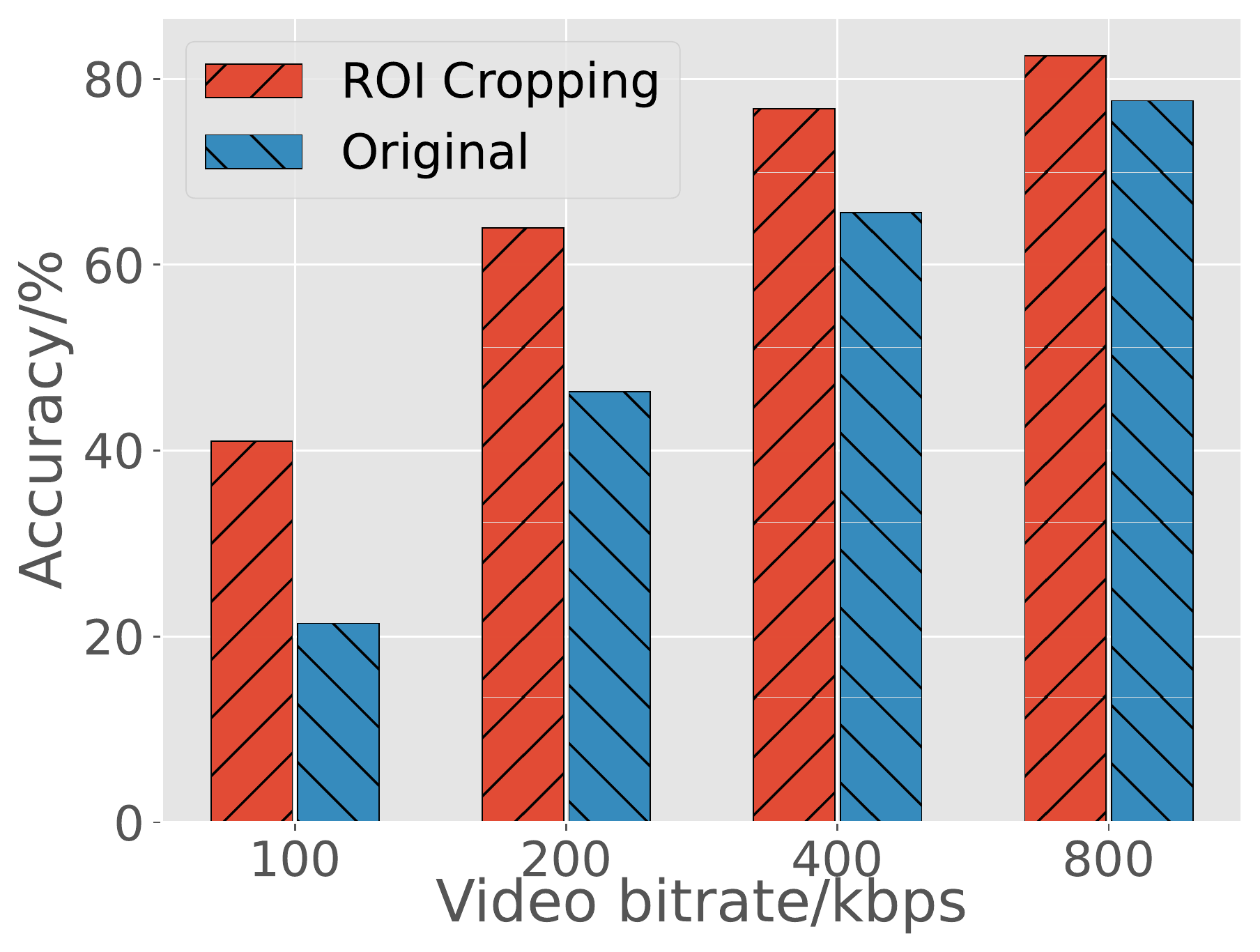}
      \caption{1080p.}
      \label{fig:acc-1080}
  \end{subfigure}
  \vspace{-3mm}
  \caption{A comparison of detection accuracy between using ROIDet and using the original frames when video segments are encoded into different resolutions (720p and 1080p) and different bitrates (100Kbps, 200Kbps, 400Kbps, and 800Kbps)}
  \label{fig:acc-diff}
 \end{figure}

\begin{figure}
\begin{minipage}[t]{0.54\linewidth}
    \includegraphics[width=\linewidth]{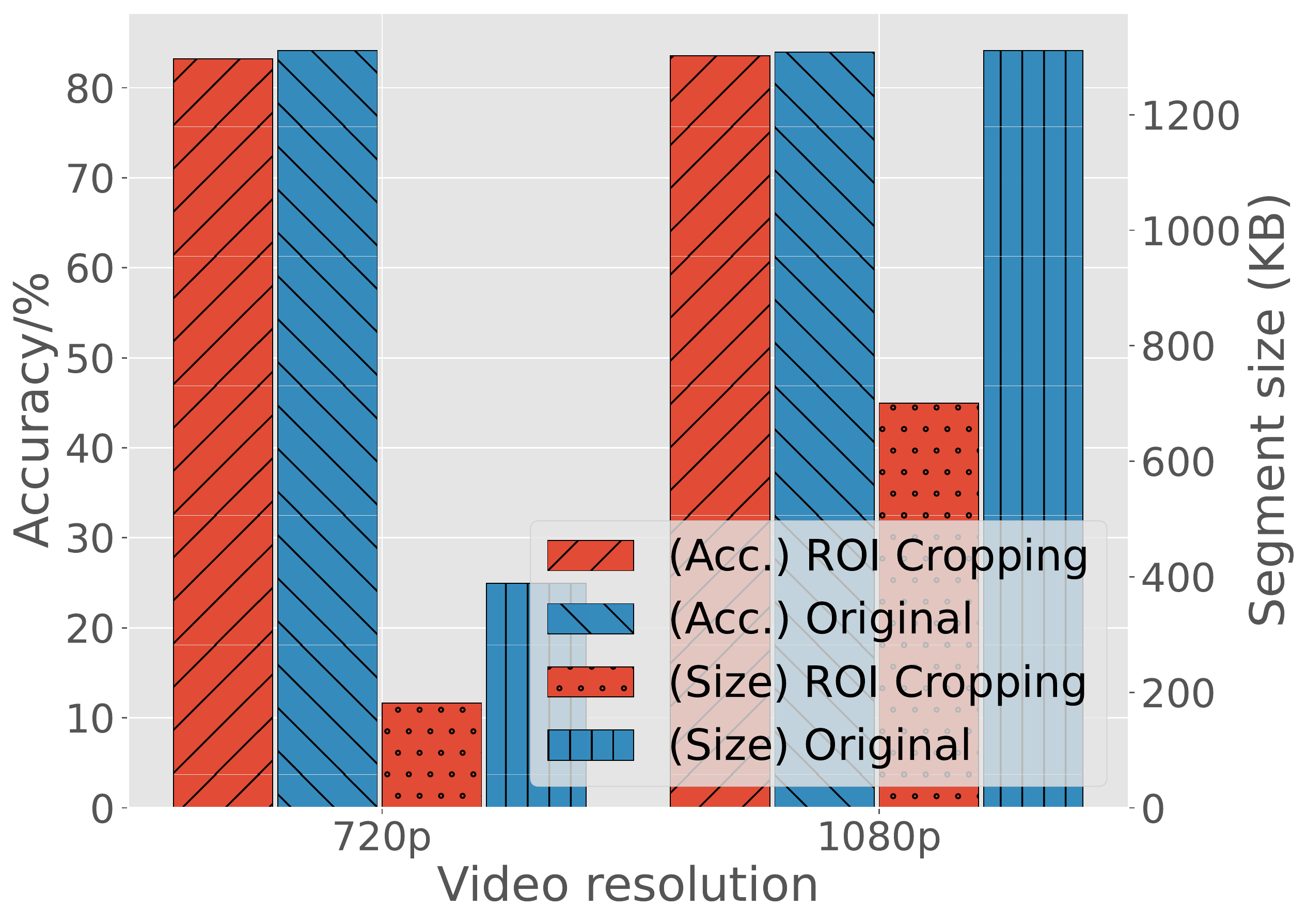}
    \caption{Compare average detection accuracy and segment size when cropped and original frames are compressed with the same CRF quality.}
    \label{fig:crf}
\end{minipage}%
    \hfill%
\begin{minipage}[t]{0.42\linewidth}
    \includegraphics[width=\linewidth]{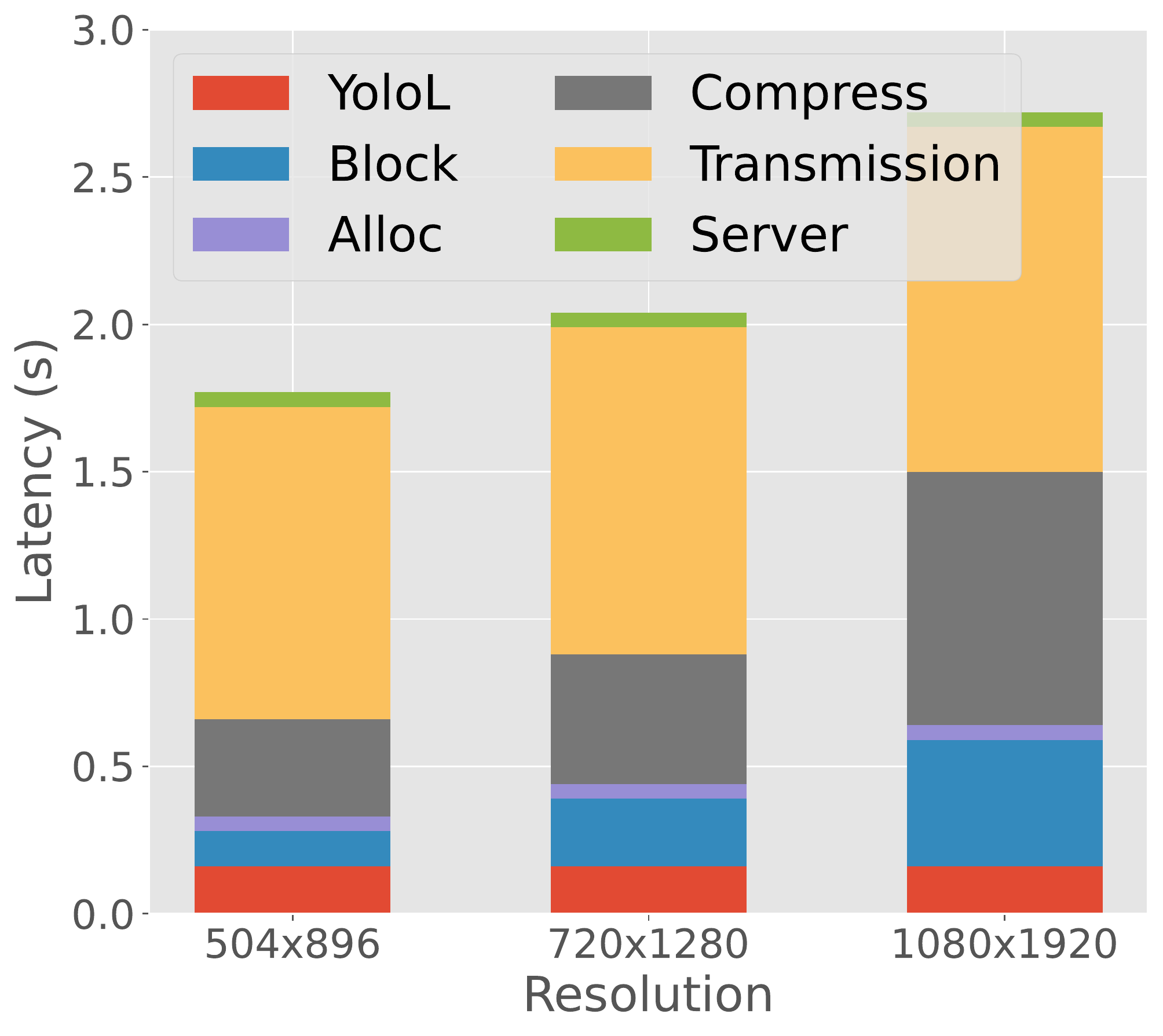}
    \caption{End-to-end latency breakdown of \name{} under different resolutions.\protect\footnotemark}
    \label{fig:bd}
\end{minipage} 
\end{figure}

The results of the \emph{first} part are shown in Figure~\ref{fig:acc-diff}. Since ROIDet crops out task irrelevant regions in video frames, more bits can be employed to compress ROIs, resulting in a higher quality compression on task related regions. Therefore, detection accuracy is boosted given the same video compression bitrate. 
\emph{Second}, we evaluate whether ROIDet sacrifices detection accuracy given sufficient compression bitrate. For this task, we compress cropped and uncropped original videos using the CRF mode. We compress with CRF value 18 for both videos which is considered visually lossless. We compare the detection accuracy of both sets of videos and the average video segment file sizes. The results are shown in Figure~\ref{fig:crf}. From the two figures we can see that ROIDet saves  $\sim 50\%$ of file sizes while inducing less than $1\%$ accuracy drop. 

\footnotetext{``YoloL'' refers to the execution time of light-weighted YOLO on camera. ``Block'' refers to the Block Based Motion Detection Algorithm. ``Alloc'' refers to bandwidth allocation algorithm, including round trip time to transmit features and decisions. ``Compress'' refers to video segment compression time. ``Transmission'' refers to the video data transmission latency. ``Server'' means the processing time on the server side.}

 \subsection{End-to-end Latency Breakdown}
 
 In this part, we perform an end-to-end analysis on \name{}'s latency. 
 We examine the latency performance of \name{} for each resolution configuration and show the results in Figure~\ref{fig:bd}. The results are obtained by averaging over all videos and all bitrates\footnote{We compute the average latency over the first four cameras since the fifth camera produces different resolutions.}. It is worth mentioning that latency is linked to video resolution, not bitrate. Bandwidth allocation adapts to available bandwidth, so transmission latency is similar for different bitrates. Note also that the processing time on the camera (``YoloL''+``Block''+``Compress'') exceeds the length of a time slot which is $1s$,  limited by the processing power of the hardware (Raspberry Pi) we use as the camera device. The camera side has a 4-core CPU, so it is able to run ROIDet and compression simultaneously for consecutive video segments.

\section{Related Work}


\textbf{Reduce Spatial or Temporal Redundancy in Videos.}
Video spatial or temporal redundancy can be leveraged in single or multiple camera streaming. Reducto\cite{li2020reducto}, NoScope\cite{kang2017noscope} and FilterForward\cite{canel2019scaling} design image features based models or light computer vision models to determine whether frames can be filtered out or not. Some works reduce task-irrelevant areas in video frames to save bandwidth for Regions of Interest using heavy computer vision models on edge or feedback regions from the server\cite{du2020server, zhang2021elf, liu2019edge, du2022accmpeg, zeng2020distream}. DeepStream has light-weighted algorithms for real-time spatial redundancy on resource-constrained devices, unlike state-of-the-art techniques that require heavy processing.
Spatula\cite{jain2020spatula} and CrossROI\cite{guo2021crossroi} study the spatial and temporal correlations across multipe cameras, but they don't perform resource allocations. 

\noindent\textbf{Resource Allocation for Video Analytic Systems.}
Other works focus on resource allocation for single or multiple video streams. JCAB\cite{wang2020joint} optimizes configurations and bandwidth for multiple cameras, but doesn't consider video spatial and temporal redundancy and their utility function doesn't consider video content. DeepDecision\cite{ran2018deepdecision} models offloading for a single camera video stream as an optimization problem. Vigil\cite{zhang2015design} and AWStream\cite{zhang2018awstream} schedules traffic for multiple co-located cameras, but they don't adapt to video content variations. VideoStorm\cite{zhang2017live} focuses on computing resources allocation on the server.  Rong \emph{et al.} propose a framework to partition DNN model execution across camera-edge-cloud infrastructure in \cite{rong2021scheduling}. DeepRT\cite{yang2021deeprt} proposes a GPU scheduling algorithm for running computer vision applications on edge.

\section{Conclusion}

In this work we propose \name{}, we propose \name{}, a bandwidth-efficient multi-camera video streaming system for deep learning analytics. It includes ROIDet for real-time removal of irrelevant regions and the Bandwidth Allocation Module for optimized detection utility. \name{} also elastically allocates bandwidth resources to adapt to correlated video content.

\bibliographystyle{ACM-Reference-Format}
\bibliography{ref}

\end{document}